\documentclass[amssymb,epsfig,color]{article}
\usepackage{epsfig}
\usepackage{amsfonts}
\usepackage{amssymb}
\usepackage{indentfirst}
\usepackage{color}
\usepackage{mathrsfs}
\def\bc{\begin{center}}

\def\ec{\end{center}}
\def\be{\begin{eqnarray}}
\def\ee{\end{eqnarray}}

\definecolor{dyellow}{rgb}{1.,0.8,.0}
\definecolor{myblue}{rgb}{.1,.1,.7}
\definecolor{dcyan}{rgb}{.0,.6,.6}
\definecolor{dmagenta}{rgb}{0.6,0.0,0.6}
\definecolor{brown}{rgb}{0.6,0.2,0.}
\definecolor{darkblue}{rgb}{.0,.0,0.5}
\definecolor{darkred}{rgb}{0.75,0.0,0.0}
\definecolor{orange}{rgb}{1.,.6,.0}
\definecolor{dorange}{rgb}{0.8,.4,.0}
\definecolor{darkgreen}{rgb}{0.0,0.6,0.0}
\definecolor{purple}{rgb}{.4,.0,.4}
\begin{document}
\baselineskip=16pt
\newcommand{\omits}[1]{}

%\hfill{\bf USTC-ICTS-07-05} \vspace{1.0cm}

\begin{center}
{\Large \bf Unparticle Physics on Cosmic Ray Photon and $e^\pm$}\\
\vspace{1cm}
Shao-Xia Chen\footnote{ruxanna@sdu.edu.cn}\\
{\em School of Space Science and Physics,}\\
 {\em Shandong University at Weihai,}\\
{\em Weihai, Shandong 264209, China}\\
Rong Hu\footnote{hur@ihep.ac.cn}\\
{\em School of Mechanical Engineering,}\\
{\em Beijing Technology and Business University,}\\
{\em Beijing 100048, China}\\
\end{center}

\begin{abstract}

We study the effects of unparticle physics on the cosmic ray photon
and $e^\pm$, including on the pair production (PP) and elastic
scattering (ES) of cosmic ray photon off various background
radiations, and on the inverse Compton scattering of cosmic ray
$e^\pm$ with cosmic radiations. We compute the spin-averaged
amplitudes squared of three processes and find that the advent of
unparticle will never significantly change the interactions of
cosmic ray photon and $e^\pm$ with various background radiations,
although the available papers show that ES which occurs in the
tree-level through unparticle exchanges will easily surpass PP in
the approximate parameter regions.

\end{abstract}

\vspace{1.0cm}


\section{Introduction}
In  convention, it is convinced that the dominant energy loss is PP
instead of ES in the Standard Model (SM) for the cosmic ray photon
with energy above the PP threshold $E_{\rm th}$ \cite{review-ray}
  \begin{equation}\label{pp-th}
   E_{\rm th}=m_e^2/\epsilon\simeq2.6\times10^{11}{\rm eV}\times
   \left(\displaystyle\frac{\epsilon}{\rm eV}\right)^{-1}~,
\end{equation}
where $\epsilon$ is the energy of a background photon. However
recent research \cite{EC1,EC2,EC3} on diphoton interaction reveals
that the cross section of unparticle exchange can easily surpass the
SM one at high enough energy because unparticle exchanges are also
at the tree-level through all $s$-, $t$-, and $u$-channels. It is
natural to explore the consequence of unparticle physics on the
cosmic ray photon, especially on whether the appearance of
unparticle will lead to its dominant energy loss process to change
from PP to ES, which will cause nontrivial observational signals in
the spectrum of cosmic ray photon.

In the meanwhile, very recently the Pamela collaboration announced
their first measurements on the cosmic ray (CR) positron fraction
\cite{p1} in the energy range $1.5-100$GeV. The positron fraction of
Pamela data shows a prominent excess to the background estimation
\cite{t1,t2} of the conventional CR propagation model in the region
$\sim10-100$GeV. This result is consistent with previous
measurements by, e.g., HEAT \cite{p3} and AMS \cite{p4}. On the
other hand, the electron spectrum up to several TeV measured by ATIC
collaboration also displays an obvious excess in the region around
$300 \sim 800 $GeV \cite{e5}, which confirms the measurements of the
electron spectrum by PPB-BETS \cite{e6}, H.E.S.S. \cite{e7,e8}, and
most recently by Fermi \cite{e9}. The mismatch between theory and
observations stimulates a lot of interest on the cosmic ray $e^\pm$,
and we will reexamine the propagation of cosmic ray $e^\pm$ in the
framework of unparticle physics. On particular, we address the
dominant loss process for cosmic ray $e^\pm$, inverse Compton
scattering, to study the impact of unparticle stuff on $e^\pm$ and
further on the observational excess.

The paper is organized as follows. In the next section, we overview
the basic property of unparticle physics, including the odd
propagator and phase space of unparticle stuff with different
Lorentz structures. In Sec.3, we derive the scattering amplitudes
for the involved processes, that is, the PP and ES for the photon
and the Compton scattering for $e^\pm$. In Sec.4, we apply the
results in the previous section to the cosmic ray physics and
analyze the specific cases to draw definite results to the
propagation of cosmic ray photon and $e^\pm$. In the final section,
we present some comments on this manuscript.
\section{Basic property of Unparticle stuff}
Two years ago, Georgi \cite{Georgi1} proposed the existence of
unparticle, which is a scale invariant sector with a non-trivial
infrared fixed-point. He assumed that the very high energy theory
contains both the SM fields and the fields of a theory with a
nontrivial infrared fixed point, which we will call ${\cal BZ}$ (for
Banks-Zaks \cite{BZ}) fields. The two sectors interact through the
exchange of particles with a large mass scale $M_{\cal U}$. Below
the scale $M_{\cal U}$, there are nonrenormalizable couplings
involving both SM fields and ${\cal BZ}$ fields suppressed by powers
of $M_{\cal U}$. These have the generic form
\begin{equation}\label{BZ-SM}
    \displaystyle\frac{1}{M_{\cal U}^k}O_{\rm sm}O_{\cal BZ}
\end{equation}
where $O_{\rm sm}$ is an operator with mass dimension $d_{\rm sm}$
built out of SM fields and $O_{\cal BZ}$ is an operator with mass
dimension $d_{\cal BZ}$ built out of ${\cal BZ}$ fields. The
renormalizable couplings of the ${\cal BZ}$ fields then cause
dimensional transmutation as scale-invariance in the ${\cal BZ}$
sector emerges at an energy scale $\Lambda_{\cal U}$. In the
effective theory below the scale $\Lambda_{\cal U}$ the ${\cal BZ}$
operators match onto unparticle operators, and the interactions of
(\ref{BZ-SM}) match onto interactions of the form
\begin{equation}\label{U-SM}
    \displaystyle\frac{C_{\cal U}\Lambda_{\cal U}^{d_{\cal BZ}-d_{\cal U}}}
    {M_{\cal U}^k}O_{\rm sm}O_{\cal U}~,
\end{equation}
where $d_{\cal U}$ is the scale dimension of the unparticle operator
$O_{\cal U}$ and the constant $C_{\cal U}$ is a coefficient
function.

It was also pointed out \cite{Georgi1} that an unparticle stuff with
scale dimension $d_{\cal U}$ looks like a non-integral number
$d_{\cal U}$ of invisible particles. In the same Letter
\cite{Georgi1}, Georgi derived the peculiar phase space of
unparticle from the scale invariance
\begin{equation}\label{phase}
    d\Phi_{\cal U}(P_{\cal U})=A_{d_{\cal U}}\theta(P_{\cal U}^0)\theta(P_{\cal U}^2)(P_{\cal
    U}^2)^{d_{\cal U}-2}~,~~~~~~~~~~A_{d_{\cal U}}=\displaystyle\frac{16\pi^{5/2}}{(2\pi)^{2d_{\cal U}}}
\displaystyle\frac{\Gamma(d_{\cal U}+1/2)}{\Gamma(d_{\cal
U}-1)\Gamma(2d_{\cal U})}~.
\end{equation}
Then, he calculated the real emission of unparticle and argued that
this kind of peculiar distribution of missing energy may be the
signal of unparticle experimentally.

Subsequently, the odd propagators were worked out independently in
\cite{Georgi2,1} for scalar, vector and tensor unparticles,
respectively,
\begin{equation}\label{propagator}
    \begin{array}{rcl}
\Delta_F(P^2)&=&Z_{d_{\cal U}}(-P^2)^{d_{\cal
U}-2}~,~~~~~~Z_{d_{\cal U}}:=\displaystyle\frac{A_{d_{\cal
U}}}{2\sin(d_{\cal
U}\pi)}\\[0.4cm]
\Delta_F(P^2)_{\mu\nu}&=&Z_{d_{\cal U}}
(-P^2)^{d_{\cal U}-2}\pi_{\mu\nu}(P)~,\\[0.4cm]
\Delta_F(P^2)_{\mu\nu,~\rho\sigma}&=&Z_{d_{\cal U}} (-P^2)^{d_{\cal
U}-2}T_{\mu\nu,~\rho\sigma}(P)~,
\end{array}
\end{equation}
where
\begin{equation}\label{note}
    \begin{array}{rcl}
(-P^2)^{d_{\cal U}-2}&=&\left\{
\begin{array}{rcl}
&&|P^2|^{d_{\cal U}-2}~~~~~~~~~~~~~~~{\rm
if}~P^2{\rm~is~negative~and~real}~, \\ [0.1cm]
&&|P^2|^{d_{\cal U}-2}e^{-id_{\cal U}\pi}~~~~~~~{\rm for~positive}~P^2{\rm~with~an~infinitesimal~}i0^\dag~, \\
\end{array}
\right. \\ [0.7cm]
\pi_{\mu\nu}(P)&=&-g_{\mu\nu}+\displaystyle\frac{P_\mu P_\nu}{P^2}~, \\[0.4cm]
T_{\mu\nu,~\rho\sigma}(P)&=& \displaystyle\frac{1}{2}\left[
\pi_{\mu\rho}(P)\pi_{\nu\sigma}(P)+\pi_{\mu\sigma}(P)\pi_{\nu\rho}(P)
-\displaystyle\frac{2}{3}\pi_{\mu\nu}(P)\pi_{\rho\sigma}(P)\right]~.
\end{array}
\end{equation}
As a direct consequence, the unusual phase in the unparticle
propagators would give rise to the interference between $s$-channel
unparticle exchange and SM amplitudes.

In this paper, we focus on the virtual exchange of unparticle at
tree-level in the interactions between the cosmic ray photon,
$e^\pm$ and background radiations, to examine the significance of
unparticle on cosmic ray photon and $e^{\pm}$.
\section{Related phenomenology}
In this section, we will derive the relevant quantities of the PP
and ES for diphoton interaction, and of the Compton scattering for
$e^\pm$. Similar processes have been examined in the previous papers
\cite{EC1,EC2,EC3,2,Compton}, however, there are several significant
distinctions in our manuscript
\begin{enumerate}
  \item There is no prior to reason to presume scalar and tensor
unparticles have the same scale dimension, $d_{\cal U}$, therefore,
we drop the subscript ${\cal U}$ which indicates unparticle stuff
and add subscript ${\rm s}$ and ${\rm t}$ to the scale dimension of
scalar and tensor unparticles to indicate their Lorentz properties,
respectively.
  \item We focus on the high energy photon and $e^\pm$,
  which permits us reasonably adopt the mass of
$e^\pm$ $m=0$.
  \item We properly write the total amplitude of an interaction as
$  i{\cal M}= i{\cal M}_{\rm SM}+i{\cal M}_{\cal U}$ and the
spin-averaged amplitude squared is given by $\overline{|{\cal
M}|^2}=\overline{{\cal M}{\cal M}^*}$, to explore the total possible
interferences.
\end{enumerate}
\subsection{Pair Production of diphoton} The diphoton PP carries on
via $t$- and $u$-channels in the SM, and the amplitude is
\begin{equation}\label{pp-sm-am}
\begin{array}{rcl}
 i{\cal M}_{\rm SM}=
 e^2\epsilon_\mu(k_1)\epsilon_\nu(k_2)
 \bar{u}(p_1)\left(\gamma^\mu
 \displaystyle\frac{i(\not\!p_1-\not\!k_1+m)}{(p_1-k_1)^2-m^2}
 \gamma^\nu+\gamma^\nu
 \displaystyle\frac{i(\not\!p_1-\not\!k_2+m)}{(p_1-k_2)^2-m^2}
 \gamma^\mu\right)v(p_2)~.
 \end{array}
\end{equation}
The diphoton PP can occur via exchanges of scalar and tensor
unparticles in $s$-channel, which gives rise to the interference
between unparticle $s$-channel and SM $t$-, $u$-channel amplitudes.
The scattering amplitudes through scalar and tensor unparticle
exchanges are
\begin{equation}\label{pp-un-am}
\begin{array}{rcl}
 i{\cal M}_{{\cal U}\rm s}&=&\displaystyle\frac{4i\lambda_0^2}
 {\Lambda_{\cal U}^{2d_{{\rm s}}-1}}\bar{u}(p_1)
 \epsilon_\mu(k_1)Z_{d_{{\rm s}}}(-s)^{d_{{\rm s}}-2}
 (k_1\cdot k_2 g^{\mu\nu}-k_1^\nu k_2^\mu)
 \epsilon_\nu(k_2)v(p_2)~, \\ [0.5cm] i{\cal M}_{{\cal U}\rm t}&=&
 \displaystyle\frac{i\lambda_2^2}{4{\Lambda_{\cal U}^{2d_{{\rm t}}}}}\bar{u}(p_1)
 [\gamma^\alpha(p_1-p_2)^\beta+\gamma^\beta(p_1-p_2)^\alpha]v(p_2)
 \epsilon_\mu(k_1)\epsilon_\nu(k_2)  Z_{d_{{\rm t}}}(-s)^{d_{{\rm t}}-2}
 \\   [0.5cm] &&
 \times T_{\alpha\beta,\rho\sigma}(k_1+k_2)
 [K^{\mu\nu\rho\sigma}(k_1^\mu,k_2^\nu)+K^{\mu\nu\sigma\rho}(k_1^\mu,k_2^\nu)]~,
 \end{array}
\end{equation}
where $T_{\mu\nu,\rho\sigma}(P)$ is defined in (\ref{note}) and
$K^{\mu\nu\rho\sigma}(p_1^\mu,p_2^\nu)$ for one photon
$\epsilon_\mu(p_1)$, the other photon $\epsilon_\nu(p_2)$ and one
tensor unparticle with Lorentz indices $\rho\sigma$ is defined as
$$
K^{\mu\nu\rho\sigma}(p_1^\mu,p_2^\nu)=-g^{\mu\nu}p_1^\rho p_2^\sigma
-p_1\cdot p_2 g^{\mu\rho}g^{\nu\sigma}+p_1^\nu p_2^\rho
g^{\mu\sigma}+p_2^\mu p_1^\rho g^{\nu\sigma}~.
$$
The spin-averaged amplitude squared is given by
\begin{equation}\label{pp-am-square}
    \overline{|{\cal M}|^2}={\bf I}+{\bf II}+{\bf III}+{\bf IV}~,
\end{equation}
where ${\bf I}$ stands for the contribution from the SM, ${\bf II}$
is that from scalar unparticle exchange, ${\bf III}$ is that from
tensor unparticle exchange, and ${\bf IV}$ is the interference
between the SM and unparticle amplitudes, respectively
\begin{equation}\label{PP-I-II-III}
    \begin{array}{rcl}
{\bf I}&=&
2e^4\left(\displaystyle\frac{t}{u}+\displaystyle\frac{u}{t}\right)~,~~~~~~~~~~~~
~~~~~~~~~~~~~~~~{\bf II}=4\lambda_0^4 Z_{d_{\rm s}}^2\left(
\displaystyle\frac{s}{\Lambda_{\cal U}^2}\right)^{2d_{\rm s}-1}~,
\\ [0.5cm] {\bf III}&=& \displaystyle\frac{\lambda_2^4 Z_{d_{\rm t}}^2}{2}\left(
\displaystyle\frac{s}{\Lambda_{\cal U}^2}\right)^{2d_{\rm
t}}\displaystyle\frac{tu}{s^2}\left(
\displaystyle\frac{t^2}{s^2}+\displaystyle\frac{u^2}{s^2}\right)~,
~~~
 {\bf IV}= 2e^2\lambda_2^2 Z_{d_{\rm t}}
\cos(d_{\rm t}\pi) \left(\displaystyle\frac{s}{\Lambda_{\cal
U}^2}\right)^{d_{\rm t}} \left(
\displaystyle\frac{t^2}{s^2}+\displaystyle\frac{u^2}{s^2}\right)~.
    \end{array}
\end{equation}
It is worth noticing that, due to the phase factor $\exp(-id_{\cal
U}\pi)$ related to the $s$-channel from the unparticle sector, there
exists interference term ${\bf IV}$ between the SM and unparticle
amplitudes which is a clear signature of unparticle physics.

The similar process $e^-+e^+\to \gamma+\gamma$ in unparticle physics
has been pursued in Ref.\cite{2},  and we can straightforwardly
obtain the spin-averaged amplitude squared for $e^-+e^+\to
\gamma+\gamma$ as
\begin{equation}\label{colider-am-square}
\begin{array}{rcl}
    \overline{|{\cal M}|^2}&=&{\bf I}+{\bf II}+{\bf III}+{\bf IV}~, \\ [0.5cm]
{\bf I}&=&2e^4\left(
\displaystyle\frac{u}{t}+\displaystyle\frac{t}{u}\right)~,~~~~~~~
~~~~~~~~~~~~~~~~ {\bf II}= \displaystyle\frac{\lambda_2^4 Z_{d_{\rm
t}}^2}{2} \left( \displaystyle\frac{s}{\Lambda_{\cal
U}^2}\right)^{2d_{\rm t}} \displaystyle\frac{tu}{s^2}\left(
\displaystyle\frac{t^2}{s^2}+\displaystyle\frac{u^2}{s^2}\right)~, \\
[0.5cm] {\bf III} &=&4\lambda_0^4Z_{d_{\rm s}}^2\left(
\displaystyle\frac{s}{\Lambda_{\cal U}^2}\right)^{2d_{\rm s}-1}~,
~~~~~~~~~~~~~~~{\bf IV}=-2e^2Z_{d_{\rm t}}\lambda_2^2\left(
\displaystyle\frac{s}{\Lambda_{\cal U}^2}\right)^{d_{\rm t}}\left(
\displaystyle\frac{t^2}{s^2}+\displaystyle\frac{u^2}{s^2}\right)
\cos(d_{\rm t}\pi)~,
\end{array}
\end{equation}
where ${\bf I}$ is from the SM, ${\bf II}$ is from the tensor
unparticle exchange, and the interference between the SM and
unparticle amplitudes is provided as ${\bf IV}$ , which are all
consistent with Ref.\cite{2}. The term ${\bf III}$ is the
contribution from the scalar unparticle exchange, which is absent in
Ref.\cite{2}. We investigate further the process in the
center-of-momentum system of the two initial photons, and the
Mandelstam variables can be written as $|t|= s(1-\cos\theta)/2$ and
$|u|= s(1+\cos\theta)/2$, and $\theta$ is the scattering angle.
Plotting the terms ${\bf I}$, ${\bf II}$, ${\bf III}$, and ${\bf
IV}$ versus $\theta$ in Fig.1, we find out that contribution ${\bf
III}$ from scalar unparticle exchange to the process $e^-+e^+\to
\gamma+\gamma$ is consequential.
\begin{center}
\includegraphics[totalheight=9.0cm]{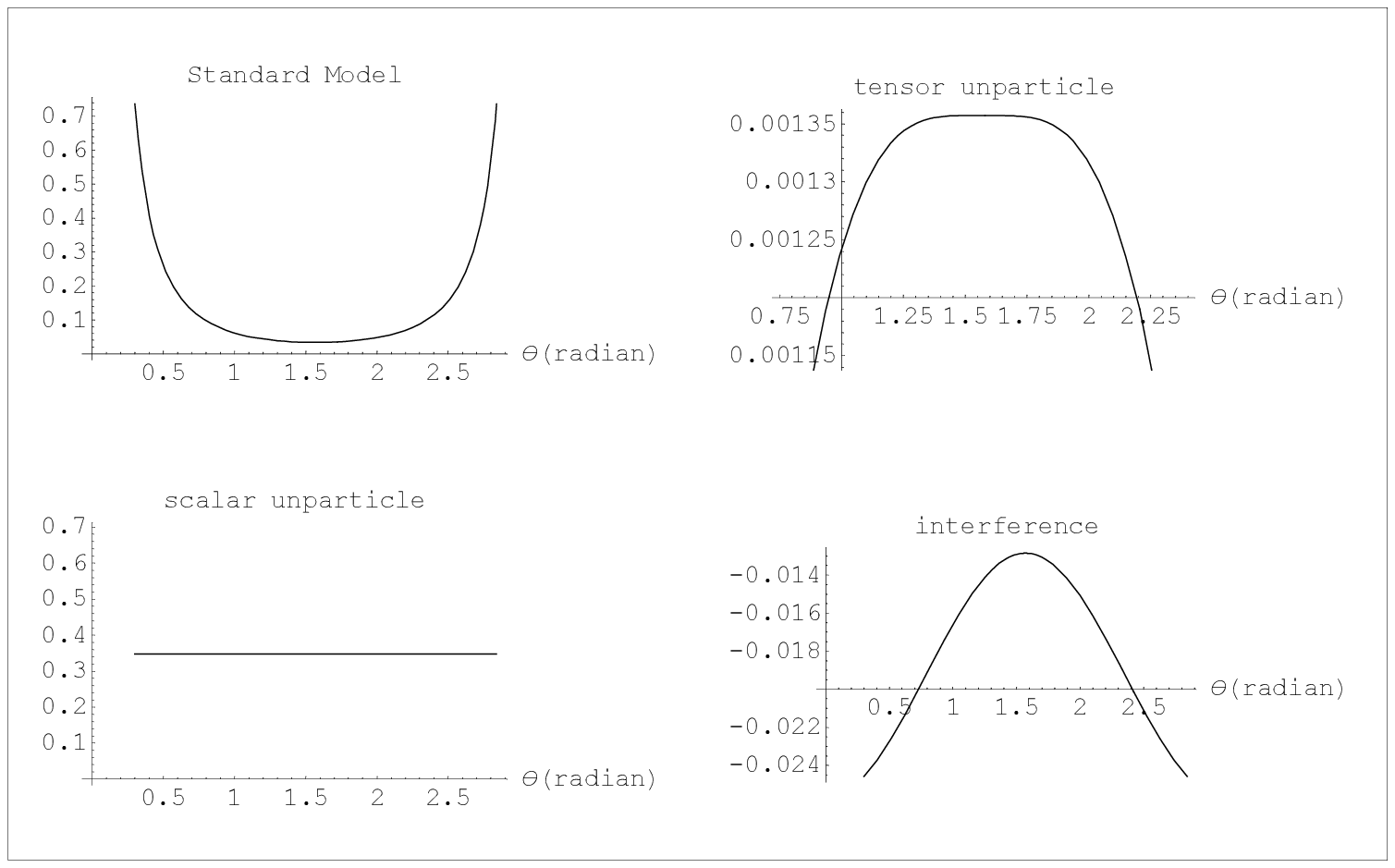}
\end{center}
{\bf Fig.1} The terms ${\bf I}$, ${\bf II}$, ${\bf III}$ and ${\bf
IV}$ in $\overline{|{\cal M}|^2}$ of $e^-+e^+\to \gamma+\gamma$
versus $\theta$ in the case $d_{\rm t}=d_{\rm s}=1.1$,
$\lambda_0=\lambda_2=1.0$ and $\Lambda_{\cal U}=1.0$TeV at
$\sqrt{s}=0.5$TeV.
\subsection{Elastic scattering of diphoton}
The diphoton can only elastically scatters via the loop-level in the
SM and thus is highly suppressed, however, it can take place via
scalar and tensor unparticle exchanges in all $s$-, $t$-, and
$u$-channels at the tree-level in the unparticle physics. The
scattering amplitude through scalar unparticle exchange is
\cite{EC1}
\begin{equation}\label{ec-scalar-M}
   \begin{array}{rcl}
     i{\cal M}_{\rm s}&=&-\displaystyle\frac{16i\lambda_0^2Z_{d_{{\rm s}}}}{\Lambda_{\cal U}^4}
     ({\cal M}_s+{\cal M}_t+{\cal M}_u)^{\mu\nu\rho\sigma}\epsilon^*_\sigma(k_1)
     \epsilon^*_\rho(k_2)\epsilon_\nu(p_1)\epsilon_\mu(p_2)~, \\ [0.5cm]
     {\cal M}_s^{\mu\nu\rho\sigma}&=& \left(\displaystyle\frac{-s}
     {\Lambda_{\cal U}^2}\right)^{d_{{\rm s}}-2}
     (-k_1\cdot k_2 g^{\rho\sigma}+k_1^\rho k_2^\sigma)
     (-p_1\cdot p_2 g^{\mu\nu}+p_1^\mu p_2^\nu)~,\\ [0.5cm]
     {\cal M}_t^{\mu\nu\rho\sigma} &=& \left(\displaystyle\frac{-t}
     {\Lambda_{\cal U}^2}\right)^{d_{{\rm s}}-2}
     (k_2\cdot p_2 g^{\mu\rho}-k_2^\mu p_2^\rho)
     (k_1\cdot p_1 g^{\nu\sigma}-k_1^\nu p_1^\sigma)~, \\ [0.5cm]
     {\cal M}_u^{\mu\nu\rho\sigma} &=& \left(\displaystyle\frac{-u}
     {\Lambda_{\cal U}^2}\right)^{d_{{\rm s}}-2}
     (k_2\cdot p_1 g^{\nu\rho}-k_2^\nu p_1^\rho)
     (k_1\cdot p_2 g^{\mu\sigma}-k_1^\mu p_2^\sigma)~.
   \end{array}
\end{equation}
The scattering amplitude through tensor unparticle exchange is
\begin{equation}\label{tensor-ec}
i{\cal M}_{{\rm t}}=i{\cal M}_{{\rm t}s}+i{\cal M}_{{\rm t}t}+i{\cal
M}_{{\rm t}u}~,
\end{equation}
$$
\begin{array}{rcl}
i{\cal M}_{{\rm t}s}&=&-\displaystyle\frac{i\lambda^2_2 Z_{d_{{\rm
t}}}}{\Lambda_{\cal U}^{2d_{{\rm t}}}}
 \times(-s)^{d_{{\rm t}}-2}T_{ab,cd}(p_1+p_2)[K^{\mu\nu ab}(p_1^\nu,p_2^\mu)+
 K^{\mu\nu ba}(p_1^\nu,p_2^\mu)][K^{\rho\sigma cd}(k_1^\sigma,k_2^\rho)+
K^{\rho\sigma dc}(k_1^\sigma,k_2^\rho)]~,
\\ [0.4cm]
i{\cal M}_{{\rm t}t}&=&-\displaystyle\frac{i\lambda^2_2 Z_{d_{{\rm
t}}}}{\Lambda_{\cal U}^{2d_{{\rm t}}}}
 \times (-t)^{d_{{\rm t}}-2}T_{ab,cd}(p_1-k_1)[K^{\nu\sigma ab}(p_1^\nu,k_1^\sigma)+
 K^{\nu\sigma ba}(p_1^\nu,k_1^\sigma)][K^{\mu\rho cd}(p_2^\mu,k_2^\rho)+
K^{\mu\rho dc}(p_2^\mu,k_2^\rho)]~,
\\ [0.4cm]
i{\cal M}_{{\rm t}u}&=&-\displaystyle\frac{i\lambda^2_2 Z_{d_{{\rm
t}}}} {\Lambda_{\cal U}^{2d_{{\rm t}}}}
 \times (-u)^{d_{{\rm t}}-2}T_{ab,cd}(p_1-k_2)[K^{\nu\rho ab}(p_1^\nu,k_2^\rho)+
 K^{\nu\rho ba}(p_1^\nu,k_2^\rho)][K^{\mu\sigma cd}(p_2^\mu,k_1^\sigma)+
K^{\mu\sigma dc}(p_2^\mu,k_1^\sigma)]~.
\end{array}
$$
The spin-averaged amplitude squared in the tree-level is given by
\begin{equation}\label{ec-am-square}
    \overline{|{\cal M}|^2}={\bf I}+{\bf II}+{\bf III}~,
\end{equation}
where ${\bf I}$ stands for the contribution from the scalar
unparticle exchange, ${\bf II}$ is that from the tensor unparticle
exchange, and ${\bf III}$ is the interference between the amplitudes
of scalar and tensor unparticle exchanges, respectively. The first
two are in good agreement with Ref.\cite{EC1}
\begin{equation}\label{ec-I-II-III}
    \begin{array}{rcl}
{\bf I}&=&\displaystyle\frac{16\lambda_0^4Z^2_{d_{\rm s}}}
    {\Lambda_{\cal U}^{4d_{\rm s}}}\left\{s^{2d_{\rm s}}+|t|^{2d_{\rm s}}
    +|u|^{2d_{\rm s}}+\cos(d_{\rm s}\pi)[(s|t|)^{d_{\rm s}}+(s|u|)^{d_{\rm s}}]
    +(|t||u|)^{d_{\rm s}}\right\}~, \\ [0.5cm] {\bf II}&=&
    \displaystyle\frac{\lambda_2^4Z^2_{d_{\rm t}}}
    {2\Lambda_{\cal U}^{4d_{\rm t}}}\left\{s^{2d_{\rm t}-4}(t^4+u^4)
    +|t|^{2d_{\rm t}-4}(s^4+u^4)+|u|^{2d_{\rm t}-4}(s^4+t^4)\right. \\ [0.4cm]
    && \left.+2\cos(d_{\rm t}\pi)s^{d_{\rm t}-2}[|t|^{d_{\rm t}-2}u^4+
    |u|^{d_{\rm t}-2}t^4]
    +2(tu)^{d_{\rm t}-2}s^4\right\}~,
\\ [0.5cm] {\bf III}&=& 4\lambda_0^2\lambda^2_2 Z_{d_{{\rm s}}} Z_{d_{{\rm
t}}}\left(\displaystyle\frac{s}{\Lambda_{\cal U}^2}\right)^{d_{{\rm
s}}+d_{{\rm t}}} \left\{
\left(\displaystyle\frac{|t|}{s}\right)^{d_{{\rm s}}+2}
\left(\displaystyle\frac{|u|}{s}\right)^{d_{{\rm t}}-2}+
\left(\displaystyle\frac{|u|}{s}\right)^{d_{{\rm s}}+2}
\left(\displaystyle\frac{|t|}{s}\right)^{d_{{\rm t}}-2} \right.  \\
[0.5cm] && \left. +\cos(d_{{\rm s}}\pi)\left[
\left(\displaystyle\frac{|t|}{s}\right)^{d_{{\rm
t}}-2}+\left(\displaystyle\frac{|u|}{s}\right)^{d_{{\rm t}}-2}
\right] +\cos(d_{{\rm t}}\pi)\left[
\left(\displaystyle\frac{|t|}{s}\right)^{d_{{\rm
s}}+2}+\left(\displaystyle\frac{|u|}{s}\right)^{d_{{\rm s}}+2}
\right] \right\}~.
    \end{array}
\end{equation}
In the Ref.\cite{EC1}, the interference ${\bf III}$ is not included,
thus, we also write the Mandelstam variables as $|t|=
s(1-\cos\theta)/2$ and $|u|= s(1+\cos\theta)/2$, and plot Fig.2 to
compare the contributions ${\bf I}$, ${\bf II}$ and ${\bf III}$.
\begin{center}
\includegraphics[totalheight=8.0cm]{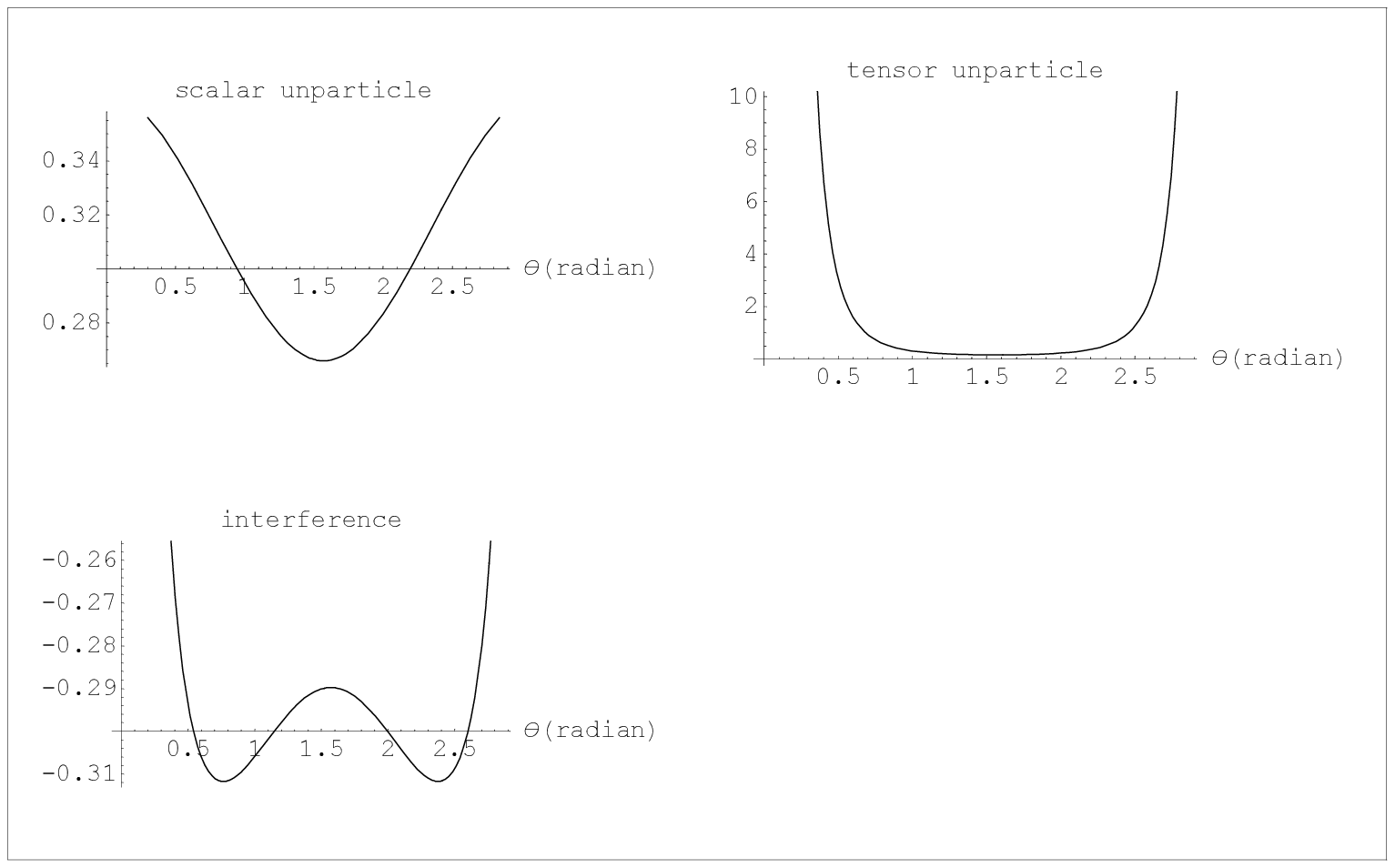}
\end{center}
{\bf Fig.2} The terms ${\bf I}$, ${\bf II}$ and ${\bf III}$ in
$\overline{|{\cal M}|^2}$ of the diphoton ES versus $\theta$ in the
case $d_{\rm t}=d_{\rm s}=1.1$, $\lambda_0=\lambda_2=1.0$ and
$\Lambda_{\cal U}=1.0$TeV at $\sqrt{s}=0.5$TeV.

\subsection{Compton scattering}
In the SM, Compton scattering can proceed via $s$- and $u$-channels,
and the scattering amplitude is
$$
i{\cal M}_{\rm SM}=-ie^2\epsilon_\mu^*(k')\epsilon_\nu(k)\bar{u}(p')
\left(\displaystyle\frac{\gamma^\mu\not\!k\gamma^\nu+2\gamma^\mu
p^\nu}{2p\cdot k}
+\displaystyle\frac{-\gamma^\nu\not\!k'\gamma^\mu+2\gamma^\nu
p^\mu}{-2p\cdot k'} \right)u(p)~.
$$
The Compton scattering through the scalar and tensor unparticle
exchanges are via $t$-channel, and the scattering amplitude is
\begin{equation}\label{cs-un}
\begin{array}{rcl}
 i{\cal M}_{{\cal U}{\rm s}}&=&
-\displaystyle\frac{4i\lambda_0^2} {\Lambda_{\cal U}^{2d_{\rm
s}-1}}\bar{u}(p')u(p)Z_{d_{\rm s}}(-t)^{d_{\rm s}-2}
 (k\cdot k'g^{\mu\nu}-k'^\nu
 k^\mu)\epsilon_\mu^*(k')\epsilon_\nu(k)~,
  \\ [0.5cm]
i{\cal M}_{{\cal U}{\rm t}}&=&\displaystyle\frac{i\lambda_2^2}
{4\Lambda_{\cal U}^{2d_{\rm
t}}}\bar{u}(p')[\gamma^\alpha(p+p')^\beta+\gamma^\beta(p+p')^\alpha]
u(p)Z_{d_{\rm t}}(-t)^{d_{\rm t}-2}T_{\alpha\beta,\rho\sigma}(p'-p)
   \\   [0.5cm] &&
 \times
(K^{\mu\nu\rho\sigma}(k^\nu,k'^\mu)+K^{\mu\nu\sigma\rho}(k^\nu,k'^\mu))
\epsilon_\mu^*(k')\epsilon_\nu(k)~.
\end{array}
\end{equation}
The spin-averaged amplitude squared has the form
\begin{equation}\label{ec-am-square}
\begin{array}{rcl}
    \overline{|{\cal M}|^2}&=&{\bf I}+{\bf II}+{\bf III}+{\bf IV}~, \\ [0.2cm]
{\bf I}&=& -2e^4 \left(
\displaystyle\frac{u}{s}+\displaystyle\frac{s}{u}\right)~,~~~~~~~~~~~~~~~~~~~~~
~~~~~~~~{\bf II} = 4\lambda_0^4 Z_{d_{\rm s}}^2 \left(
\displaystyle\frac{|t|}{\Lambda_{\cal U}^2}\right)^{2d_{\rm s}-1}~, \\
[0.5cm] {\bf III}&=&-\displaystyle\frac{\lambda_2^4Z_{d_{\rm
t}}^2}{2} \left(\displaystyle\frac{|t|}{\Lambda_{\cal
U}^2}\right)^{2d_{\rm t}} \displaystyle\frac{us}{t^2} \left(
\displaystyle\frac{u^2}{t^2}+\displaystyle\frac{s^2}{t^2}\right)
~,~~~~{\bf IV} = 2e^2\lambda_2^2 Z_{d_{\rm t}} \left(
\displaystyle\frac{|t|}{\Lambda_{\cal U}^2}\right)^{d_{\rm t}}
\left(
\displaystyle\frac{u^2}{t^2}+\displaystyle\frac{s^2}{t^2}\right)~,
\end{array}
\end{equation}
where ${\bf I}$ is the SM contribution, ${\bf II}$ is the
contribution from the scalar unparticle exchange, ${\bf III}$ is
that from tensor unparticle exchange, and ${\bf IV}$ is the
interference between the amplitudes of SM and unparticle stuff.

Ref.\cite{Compton} derived the contribution from scalar unparticle
exchange to the Compton scattering, and we complete the derivation
by replenishing ${\bf III}$ and ${\bf IV}$. Similarly we write the
Mandelstam variables as $|t|= s(1-\cos\theta)/2$ and $|u|=
s(1+\cos\theta)/2$, then contrast the contributions ${\bf I}$, ${\bf
II}$, ${\bf III}$, and ${\bf IV}$ in the Fig.3, and it is clear that
the terms ${\bf II}$ and ${\bf IV}$ are vital and innegligible.
\begin{center}
\includegraphics[totalheight=7.0cm]{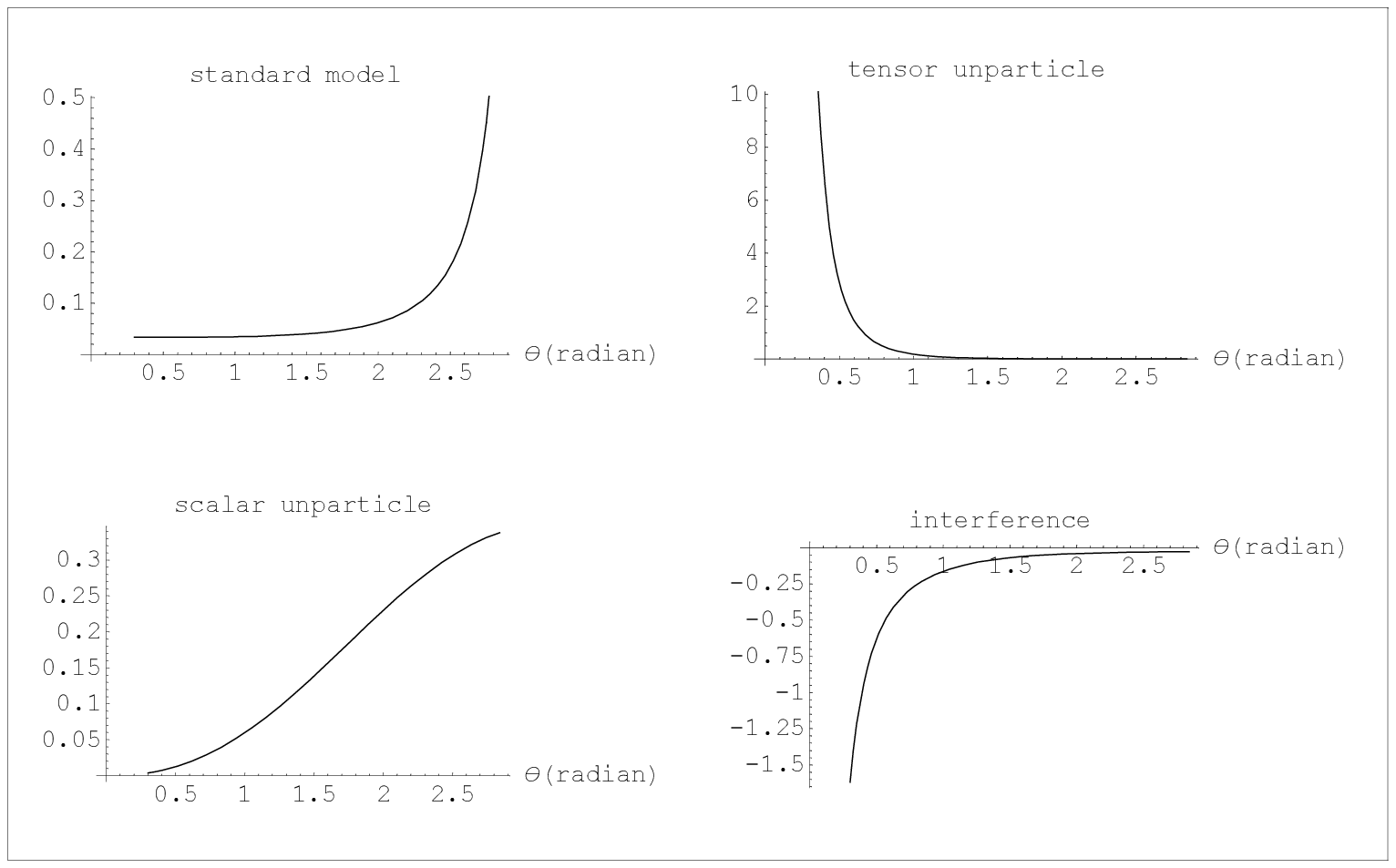}
\end{center}
{\bf Fig.3} The terms ${\bf I}$, ${\bf II}$, ${\bf III}$ and ${\bf
IV}$ in $\overline{|{\cal M}|^2}$ of the Compton scattering versus
$\theta$ in the case $d_{\rm t}=d_{\rm s}=1.1$,
$\lambda_0=\lambda_2=1.0$ and $\Lambda_{\cal U}=1.0$TeV at
$\sqrt{s}=200$GeV.

\subsection{Brief Summary}
In the SM, the ES cross section of photon-photon interaction is
nearly negligible compared to the PP one for the reason that ES can
only proceed at loop-level while PP can take place at tree-level.
However, previous analysis indicate that ES can also proceed at
tree-level in the presence of unparticle physics. In order to
contrast the probabilities of ES and PP for interacting diphoton, we
plot their $\overline{|{\cal M}|^2}~$s in Fig.4. It is evident that
in the framework of unparticle physics ES can easily exceed PP in
some parameter regions, such as that in Fig.4.
\begin{center}
\includegraphics[totalheight=3.0cm]{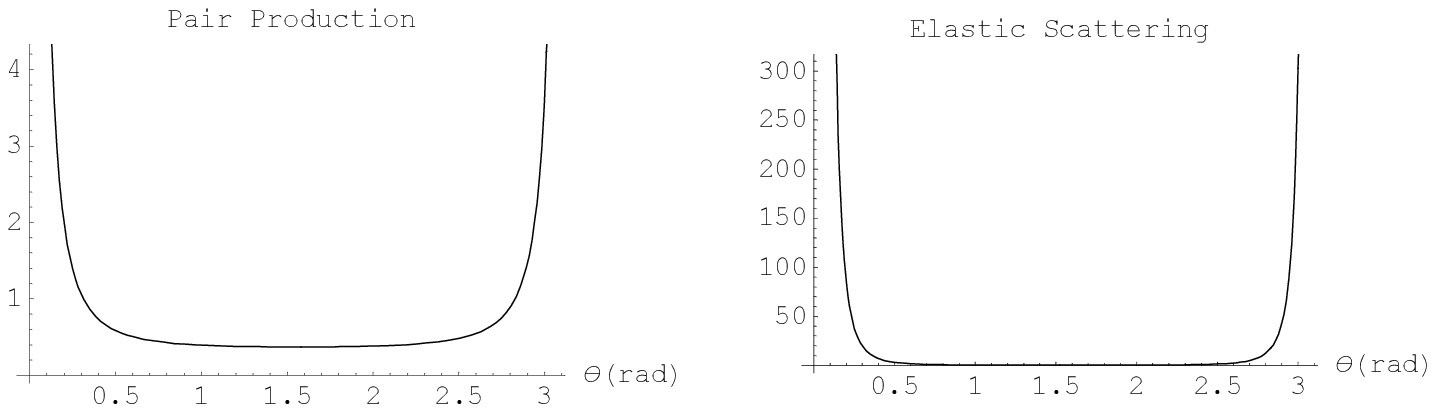}
\end{center}
{\bf Fig.4} The $\overline{|{\cal M}|^2}~$s of ES and PP of
interacting diphoton versus $\theta$ in the case $d_{\rm t}=d_{\rm
s}=1.1$, $\lambda_0=\lambda_2=1.0$ and $\Lambda_{\cal U}=1.0$TeV at
$\sqrt{s}=0.5$TeV.

\vspace{0.3cm}

In addition, as shown in Fig.5, while $d_{\rm t}=d_{\rm s}=d$
increases, the $\overline{|{\cal M}|^2}~$s of the ES, PP and Compton
scattering all have a sharp decline. Moreover, the decrease of PP
$\overline{|{\cal M}|^2}$ is much slower than Compton scattering
one, which is also much slower than ES one, with the increase of
scale dimension $d$.
\begin{center}
\includegraphics[totalheight=9.0cm]{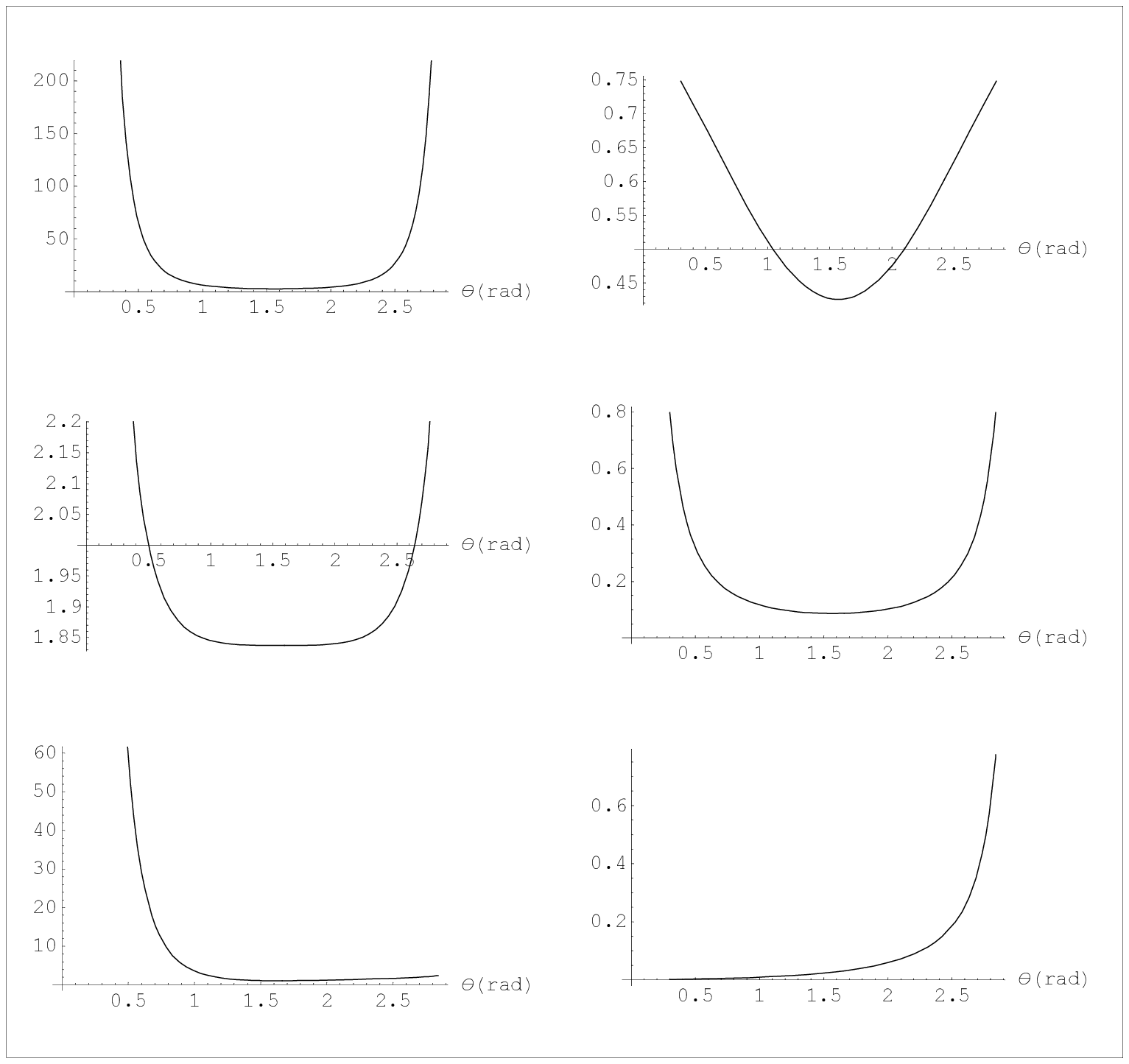}
\end{center}
{\bf Fig.5} The $\overline{|{\cal M}|^2}~$s of ES and PP for
interacting diphoton, and of the Compton scattering versus $\theta$
in the case $d_{\rm t}=d_{\rm s}=d$, $\lambda_0=\lambda_2=1.0$ and
$\Lambda_{\cal U}=1.0$TeV at $\sqrt{s}=1.0$TeV . The left column
displays those with $d=1.1$ and the right one does those with
$d=1.9$; from top to bottom, every row deals with that of ES, PP,
and Compton scattering, respectively.

\section{Unparticle physics on cosmic ray photon and $e^\pm$}
In this section, we specially investigate the previous processes in
the cosmic ray physics, that is, the ES, PP and inverse Compton
scattering in the energy scope of relevance to the cosmic ray photon
and $e^\pm$ interacting with various background radiations. As is
well known, ultra high energy cosmic rays are the highest energy
events we have observed in our earth laboratory frame. However, it
is worth clarifying that, because the typical energies of background
radiations are tiny, about or below $\sim$eV, the invariant
$\sqrt{s}$s in the interactions between cosmic rays, even for the
highest energy cosmic ray event, and various background radiations
are small compared to those related to the colliders. Due to the
different $s$ in collider and cosmic ray physics, we will prudently
deal with the previous processes in the cosmic ray physics instead
of applying the previous results immediately.

\begin{center}
\includegraphics[totalheight=13.5cm]{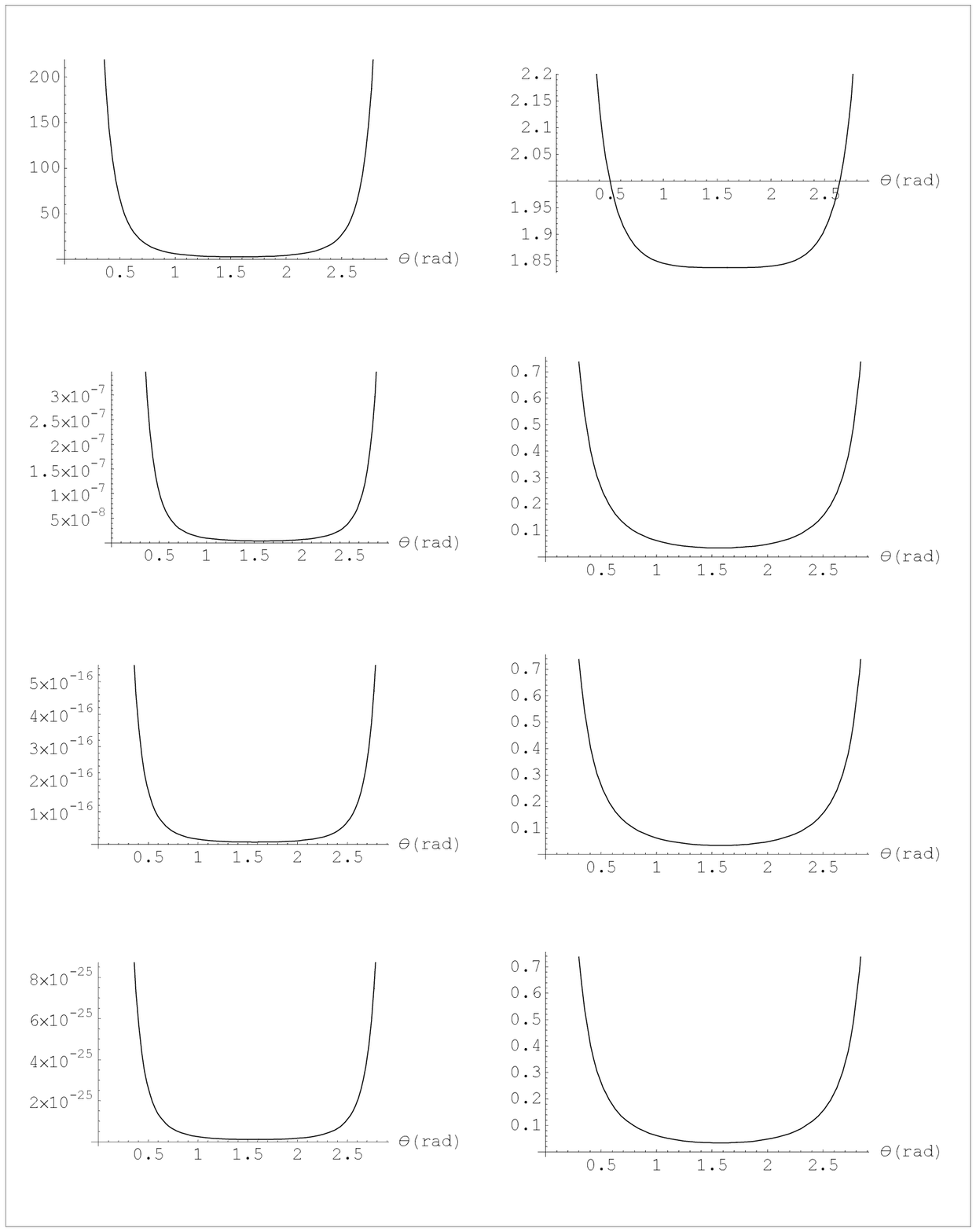}
\end{center}
{\bf Fig.6} The $\overline{|{\cal M}|^2}~$s of ES and PP for
interacting diphoton versus $\theta$ in the case $d_{\rm t}=d_{\rm
s}=1.1$, $\lambda_0=\lambda_2=1.0$ and $\Lambda_{\cal U}=1.0$TeV.
The left column exhibits ES $\overline{|{\cal M}|^2}~$s and the
right one does PP ones; from top to bottom, every row is that at
$\sqrt{s}=1000$GeV, $\sqrt{s}=10$GeV, $\sqrt{s}=0.1$GeV, and
$\sqrt{s}=10^{-3}$GeV, respectively.
\subsection{Cosmic ray photon}
The cosmic ray photon spectrum can extend from $10^8$eV to the
highest $10^{21}$eV and the typical energies $\epsilon$ of different
background radiations vary from $10^{-8}$eV to $10^{-1}$eV. Thus,
roughly speaking, the variable $\sqrt{s}=\sqrt{4E\epsilon}$ of the
interaction between a cosmic ray photon and a background photon is
in the range $(2.0{\rm eV},~20{\rm GeV})$ where $E$ is energy of the
cosmic ray photon. Allowing for the existence of threshold $E_{\rm
th}$ for PP, we plot Fig.6 to describe the variation of the
$\overline{|{\cal M}|^2}~$s of ES and PP for interacting diphoton

Fig.6 shows that the contributions of unparticle exchange to ES and
PP decrease quickly with decrease of $\sqrt{s}$. As is discussed
below Fig.4, it is obvious that in some regions of parameter space,
that is, $\sqrt{s}$ is $\sim$TeV and $d_{\rm t}$, $d_{\rm s}$ are
near above 1.0, ES is really dominant on PP. However, for cosmic ray
photons interacting with various background radiations, the $s$s are
much smaller than TeV scale, which makes the probability of ES much
smaller than PP one as shown in Fig.6. In result, in the case of
cosmic ray photon propagation, unparticle physics plays a minute
role and the dominant energy loss process of cosmic ray photon with
energy above PP threshold $E_{\rm th}$ will never convert into ES.

\subsection{Cosmic ray $e^\pm$}
Similarly $\sqrt{s}\sim\sqrt{4E\epsilon}$ of inverse Compton
scattering between cosmic ray $e^\pm$ and various background
radiations are also in the rough range $(2.0{\rm eV},~20{\rm GeV})$
where $E$ is the energy of cosmic ray $e^\pm$. We plot Fig.7 to
describe the variation of the $\overline{|{\cal M}|^2}$ of inverse
Compton scattering with variable $\sqrt{s}$.
\begin{center}
\includegraphics[totalheight=7.0cm]{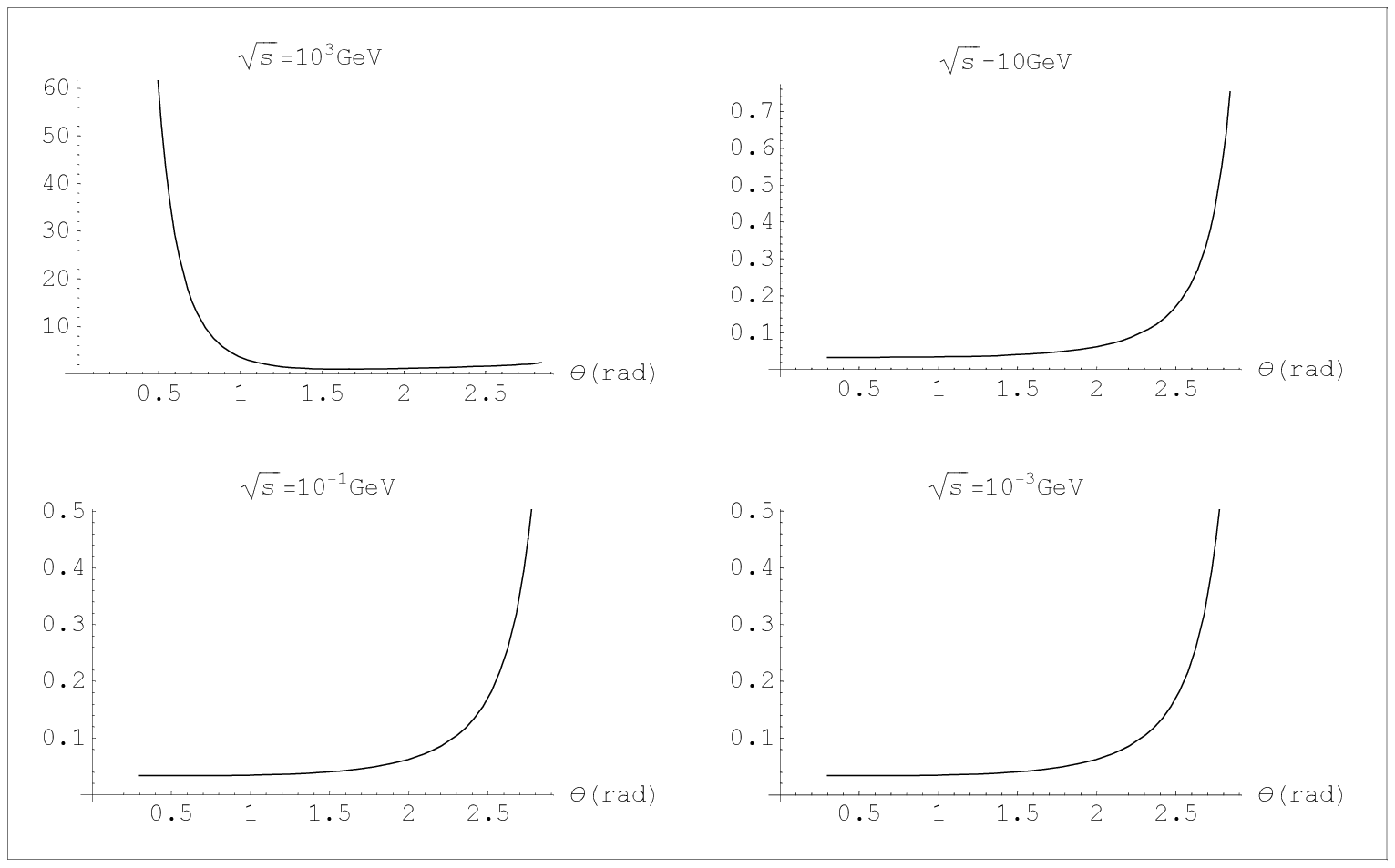}
\end{center}
{\bf Fig.7} The $\overline{|{\cal M}|^2}$ of the inverse Compton
scattering versus $\theta$ in the case $d_{\rm t}=d_{\rm s}=1.1$,
$\lambda_0=\lambda_2=1.0$ and $\Lambda_{\cal U}=1.0$TeV with
different $\sqrt{s}$.

\vspace{0.3cm}

Similar to the case of cosmic ray photon, $s$ of the inverse Compton
scattering of a cosmic ray $e^\pm$ is also small compared to those
in the colliders, which results in the impact of unparticle physics
on cosmic ray $e^\pm$ is almost negligible as Fig.7 indicates.

Now let us turn to the hot topic, observed cosmic ray $e^\pm$ excess
about the energy $100$GeV, where cosmic ray $e^\pm$ mainly interacts
with optical background radiation of energy $\sim$eV. We can obtain
figures similar to that at the bottom right corner in Fig.7, so, the
same conclusion can be drawn for cosmic ray $e^\pm$ excess: the
influence of unparticle physics on cosmic ray $e^\pm$ excess can
nearly be neglected.

In fact, as have been pointed out in papers \cite{higgs, scale}, the
low energy experiments will never be able to observe unparticle
physics. The coupling $\sim H^2O_{\cal U}$ between scalar unparticle
operator and SM Higgs boson will cause the breaking of conformal
symmetry of unparticle sector at some scale $\Lambda_{\not\!{\cal
U}}$, thus, the experimental probes of the conformal hidden sector
must probe energies in the conformal window $\Lambda_{\not\!{\cal
U}}< \sqrt{s}_{\rm exp}<\Lambda_{\cal U}$. In general,
$\Lambda_{\not\!{\cal U}}$ has the scale $\gtrsim10$GeV, and below
this scale the unparticle sector becomes a traditional particle
sector. Comparably, the typical energy
$\sqrt{s}=(\sim)\sqrt{4E\epsilon}$ of interaction between a cosmic
ray photon ($e^{\pm}$) and a background photon is in the scope
$(2.0{\rm eV},~ 20{\rm GeV})$. The incoming photon ($e^{\pm}$)
couples unparticle operators if and only if it satisfies
\begin{equation}\label{condition}
\sqrt{4E\epsilon}>\Lambda_{\not\!{\cal U}}=
\left[\left(\displaystyle\frac{\Lambda_{\cal U}}{M_{\cal
U}}\right)^{d_{\cal BZ}-d_{\cal U}} M_{\cal U}^{2-d_{\cal
U}}v^2\right]^{\frac{1}{4-d_{\cal U}}}~,
\end{equation}
where $v=\langle H\rangle=$246GeV is the vacuum expectation value
(VEV) of Higgs boson \cite{higgs}. Deviating from the kinematic
conditions (\ref{condition}) for unparticle exchange, cosmic ray
photon ($e^{\pm}$) decouples rapidly from unparticle sector, which
makes unparticle physics inaccessible for most cosmic ray events. In
the particular case of observed cosmic ray $e^\pm$ excess,
$\sqrt{s}\sim\sqrt{4E\epsilon}\sim10^6$eV is far below the
characteristic scale $\Lambda_{\not\!\cal U}\gtrsim10$GeV, thus,
unparticle physics should be irrelevant to the issue. The
conclusions drawn here more precisely qualify the above negative
results obtained from Fig.6 and Fig.7.

\section{Results and Comments}
We compute the amplitudes of PP and ES for diphoton interaction and
the amplitude of inverse Compton scattering in the framework of
unparticle physics, and we find that unparticle physics plays a
negligible role in the cosmic ray photon and $e^\pm$ propagation.

Let us close with several comments.
\begin{itemize}
  \item In some regions of parameter space of interacting diphoton,
  ES will dominate PP while in some other
  regions PP is dominant on ES, which though has a trivial influence
  on the cosmic ray physics but may play roles to
  different extents in other photon phenomena, such
  as the gamma-ray bursts and supernovae, etc.
  \item In the discussion, we firstly set the
  scalar and tensor unparticles have the different scale dimensions
$d_{\rm s}$ and $d_{\rm
  t}$. In the following, when plotting the figures we adopted $d_{\rm s}=d_{\rm
  t}$ in order to simplify the case and have a general but rough
  results. In fact, there
  seemingly exists no reason to impose $d_{\rm t}=d_{\rm s}$
  except simplicity. However, due to the crucial influence of $d_{\cal U}$ on the
$\overline{|{\cal M}|^2}~$s of the three processes, there will
appear interesting but more complicated signals in the case $d_{\rm
t}\neq d_{\rm s}$.
\item The advent of unparticle physics results in new angular
distributions in the $\overline{|{\cal M}|^2}~$s of three processes
discussed above and will further give rise to the angular
distributions in the cross sections which are very different from
those in the SM cross sections, which is a distinct signal in the
related phenomena, such as $e^+e^-$ collider.
\end{itemize}
\section*{Acknowledgments}
The work is supported in part by the Science Foundation of Shandong
University at Weihai under Grant No. 0000507300020.

%%%%%%%%%%%%%%%%%%%%%%%%%%%%%%%%%%%%%%%%%%%%%%%%%%%%%%%%%%%%%%%%%%%%%%%%%%

%\end{CJK*}

\begin{thebibliography}{999}
%%%%%%%%%%%%%%%%%%%cosmic ray
\bibitem{review-ray} P. Bhattacharjee and G. Sigl,
Phys. Rept. {\bf 327}, (2000) 109.

%%%%%%%%%%%%%%%EC
\bibitem{EC1} C. F. Chang, K. Cheung and T. C. Yuan, Phys. Lett. B {\bf 664}, (2008)
291.

\bibitem{EC2} O. Cakir, K. O. Ozansoy, Eur. Phys. J. C {\bf 56}, (2008) 279.

\bibitem{EC3} T. Kikuchi, N. Okada and M. Takeuchi, Phys. Rev. D {\bf 77}, (2008)
094012.

%%Pamela
\bibitem{p1} O. Adriani et al., arXiv:0810.4995 [astro-ph].

\bibitem{t1} I. V. Moskalenko and A. W. Strong, Astrophys. J. 493,
694 (1998).

\bibitem{t2} E. A. Baltz and J. Edsjo, Phys. Rev. D 59, 023511 (1999).

\bibitem{p3} S. W. Barwick et al. [HEAT Collaboration], Astrophys. J. 482,
L191 (1997).

\bibitem{p4} M. Aguilar et al. [AMS-01 Collaboration], Phys. Lett. B 646, 145
(2007).

%%%%%%%%%%%%%%%%%%%%%e
\bibitem{e5} J. Chang et al., Nature 456 (2008) 362.

\bibitem{e6} S. Torii et al.[PPB-BETS Collaboration], arXiv: 0809.0760.

\bibitem{e7} F. Aharonian et al. [H. E. S. S.

Collaboration], Phys. Rev. Lett. {\bf 101}, (2008) 261104.

\bibitem{e8} F. Aharonian et al. [H.E. S. S. Collaboration], arXiv: 0905.0105.

\bibitem{e9} Fermi Collaboration, Phys. Rev. Lett. {\bf 102}, (2009) 181101.

%%%%%%%%%%%%%%%%%%%%%%%%%%%%unparticle

\bibitem{Georgi1} H. Georgi, Phys. Rev. Lett. {\bf 98}, (2007) 221601.

\bibitem{BZ} T. Banks and A. Zaks, Nucl. Phys. B {\bf 196}, (1982) 189.

\bibitem{Georgi2} H. Georgi, Phys. Lett. B {\bf 650}, (2007) 275.

\bibitem{1} K. Cheung, W. Y. Keung and T. C. Yuan, Phys. Rev. Lett. {\bf 99}, (2007)
051803.

\bibitem{2} K. Cheung, W. Y. Keung and T. C. Yuan, Phys. Rev. D {\bf 76},
(2007) 055003.

%\bibitem{SM-U} S. L. Chen, X. G. He, Phys. Rev. D {\bf 76}: 091702(R), 2007

%%%%%%%%%%%%%%%%%%%%%Compton
\bibitem{Compton}  O. Cakir and K. O. Ozansoy, Europhys. Lett. {\bf 83}, (2008) 51001.

%%%%%%%%%%%%%%%%%%%%%%%%%%%%scale and Higgs
\bibitem{higgs} P. J. Fox, A. Rajaraman and Y. Shirman, Phys. Rev. D {\bf 76},
(2007) 075004.

\bibitem{scale} M. Bander, J. L. Feng, A. Rajaraman, and Y. Shirman, Phys. Rev. D {\bf 76},
(2007) 115002.


\end{thebibliography}
\end{document}